# Quantization of Black Holes in the Shielded Strong Gravity Scenario (I. Neutral Scalar States)

D. G. Coyne  (SCIPP, University of California at Santa Cruz)
D. C. Cheng  (Almaden Research Center, San Jose, CA)

A previously used quantization mechanism is applied to the continuous states of the shielded strong gravity scenario [1], yielding two types of spectra for uncharged black hole scalars.  Each yields the general morphology for states expected in this scenario at LHC and at arbitrarily higher energies, once the parameters are determined by the two lowest-lying scalar states.  A particularized example for the preferred type of quantization is numerically evaluated.

**Section 1.  Motivation**

The shielded strong gravity scenario (SSGS) was introduced in an effort to simulate what happens to the black hole evaporation process if gravity is really a very strong force but is substantially shielded [1]. It was found that a simple *ansatz* for the dependence of this type of force on energy scale (black hole temperature) regularized Hawking evaporation, eliminating its infinities and allowing extrapolation of the evaporation process into the sub-Planckian region of black hole masses (fig. 1).  Near the Planck mass, after a sudden change in behavior mimicking a phase transition, the evaporating hole assumes normal thermodynamic behavior (with positive heat capacity) and shows effective mass decay times suggestive of those of strongly-interacting elementary particles having the same mass.

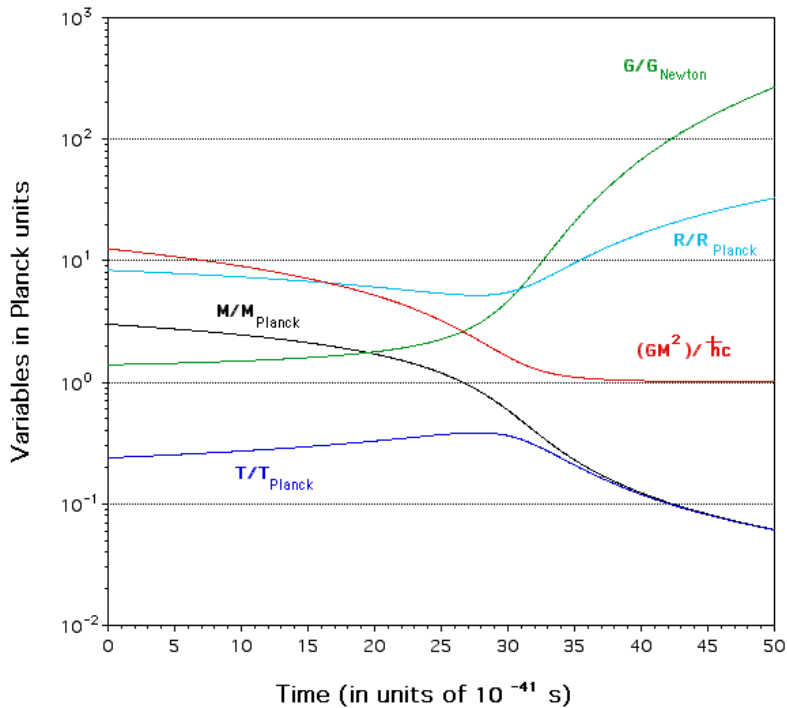

*Fig. 1:  Behavior of the variables relative to their Planck values.*

This model also included a novel generalization of the entropy of black holes. While agreeing with conventional treatments for masses greater than the Planck mass $M_P$, the sub-Planckian entropy of Schwarzschild holes is basically logarithmic as a function of mass (as will be displayed in eq. (11) and in fig. 2). This model thus has the unique property of producing the expected entropy of quantum fields at low energy (logarithmic), that of quantum strings at intermediate energies (linear), and that of black holes at energies above $M_P$ (quadratic). Appendix B of [1] strongly suggests that if elementary particles are taken to be sub-Planckian black holes with such intrinsic entropy, then black hole evaporation everywhere satisfies the second law of thermodynamics for that intrinsic entropy alone. The anomalies usually associated with elementary particles as black holes are absent in this model because the horizon area of evaporating holes is dual (and symmetrical) under the transformation $M/M_P \to M_P/M$, and this is sufficient to maintain compatibility with the uncertainty principle. (The curve $GM^2/\hbar c$, shown in fig. 1, converges to unity, implying that the horizon radius $2GM/c^2$ is always greater than the Compton wavelength $\hbar/Mc$ in all regions of $M$.)

It was speculated in [1] that some mechanism might exist that would stabilize the evaporating Schwarzschild hole at discrete values of mass, corresponding to (yet-undiscovered) neutral scalar elementary particles; no mechanism, other than the requisite thermodynamic equilibrium with the cosmic microwave background, at a current epoch mass of 6 milli-eV, was offered.

In this paper, we examine such a mechanism. Section 2 points out some previous uses of this mechanism, while Section 3 applies it to the SSGS. In Section 4 we present an overview of the results and quantitative numerical predictions for laboratory tests of the spectrum of uncharged scalars. A preview of the implications for CERN LHC experiments may be seen near the end of this paper, in fig. 7. Throughout this paper, all physical constants are explicitly displayed in order to avoid confusing the Newtonian gravitational coupling constant $G_N$ with the effective coupling $G$ – a variable in SSGS.

**Section 2. Mechanism**

Quantization of black holes with masses > $M_P$ is not a new idea, nor is quantization of the scalar states of elementary particles expected from SUSY, SUGRA, or from various forms of string theory. Black hole quantization has proceeded via identification of a minimal quantum of area, related to the area of the horizon acting as an adiabatic invariant [2, 3, 4]. Given the classical dependence of black hole area $\propto M^2$, the prediction directly follows that the black hole energy levels will be quantized with spacings $\propto 1/M$. In loop quantum gravity, quantization of area is a fundamental outcome of the theory [5,6]. In addition, black hole entropy is also proportional to the horizon area in this formulation. These two results imply a quantized black hole mass spectrum that is also $\propto 1/M$. These types of treatments usually imply a quantum of area $A_o$ that is just a multiple (of order 1) of the Planck area $A_P = \hbar G_N/c^3$. In particular, for uniformity in this paper, we use the common result of [2, 3, 4, 7]:

(1) $$A_o = 4(\ln 2)\hbar G_N/c^3$$

Another treatment originating from gravity physics [8, 9] is to associate the classical normal-mode vibrations of the black hole as corresponding to the frequencies of quanta arising from transitions between bound states (Bohr correspondence); again, this suggests a quantum as above, except ln2 in eq. (1) is replaced by ln3 (spin 1/2 networks replaced by spin 1 networks, as explained in [10]).

The other region of mass that has been subject to quantization is sub-Planckian, where now the operative theories are those of elementary particle physics. Quantization and spacing of sub-Planckian scalar states has been the conceptual playground of unification schemes and string theory, and many different putative spectra have been predicted [11]. Especially notable (because of some contrasts to the results to be found in this paper) are the predictions of "towers" of evenly spaced states [12, 13].

In this paper, we attempt to quantize not only the super-Planckian mass states of Schwarzschild black holes, but, following the SSGS, the sub-Planckian states as well, identifying the latter with the uncharged scalar particle spectrum. The quantization of the super-Planckian states will be trivially different than the Bekenstein, et al. treatment. But for the sub-Planckian states, not addressed by (or even permitted in) the Bekenstein work, we will show that there are two conceptually different approaches in SSGS. Only one of these approaches has results reasonably consistent with existing data on scalar elementary particles, e.g., their absence at currently accessible mass scales.

**Section 3.  Quantization in the SSGS**

The question in applying the Bekenstein quantization technique is this: Are we truly quantizing the physical area of a black hole, or are we quantizing the unseen information (intrinsic entropy) contained on or within its horizon? Both long ago and more recently [14], J. A. Wheeler argued persuasively that *information* is the fundamental reality of the world. Well above the Planck mass the entire question is somewhat moot: one is mathematically equivalent to the other, because the entropy

(2) $\qquad S = A/4$  (if the area $A$ is in Planck units $\hbar G_N / c^3$) .

With our above choice for $A_0$ the quantum of entropy $S_0$ would then be

(3) $\qquad S_0 = \ln 2,$  where entropy is in units of "nats",

and the number of microstates corresponding to this quantum of entropy is $\exp(S_0) = 2^1$; the entropy is one bit.

To see if the equivalence of $S$ and $A/4$ is still true in the SSGS, we will need to recall the exact forms for $M(t)$, $A(M)$, $G(M)$ and $S(M)$ from [1], expressed in units of Planck mass, Planck area, $G_N$ and nats, respectively. These complete formulae are listed in Appendix A, but their algebraic complexity obscures what the quantization process yields quantitatively. It turns out that these quantities have very simple asymptotic forms in all mass regions except for the immediate vicinity of the traditional Planck mass. Most of the quantitative results can be obtained using these forms, also given in Appendix A. The numerical details of the states will be deferred to Sec. 4.

A preliminary overview revealing the qualitative simplicity of the process and the *non-equivalence* of quantization of $S$ and $A/4$ can be seen in fig. 2, where we have used the exact forms for entropy and area.

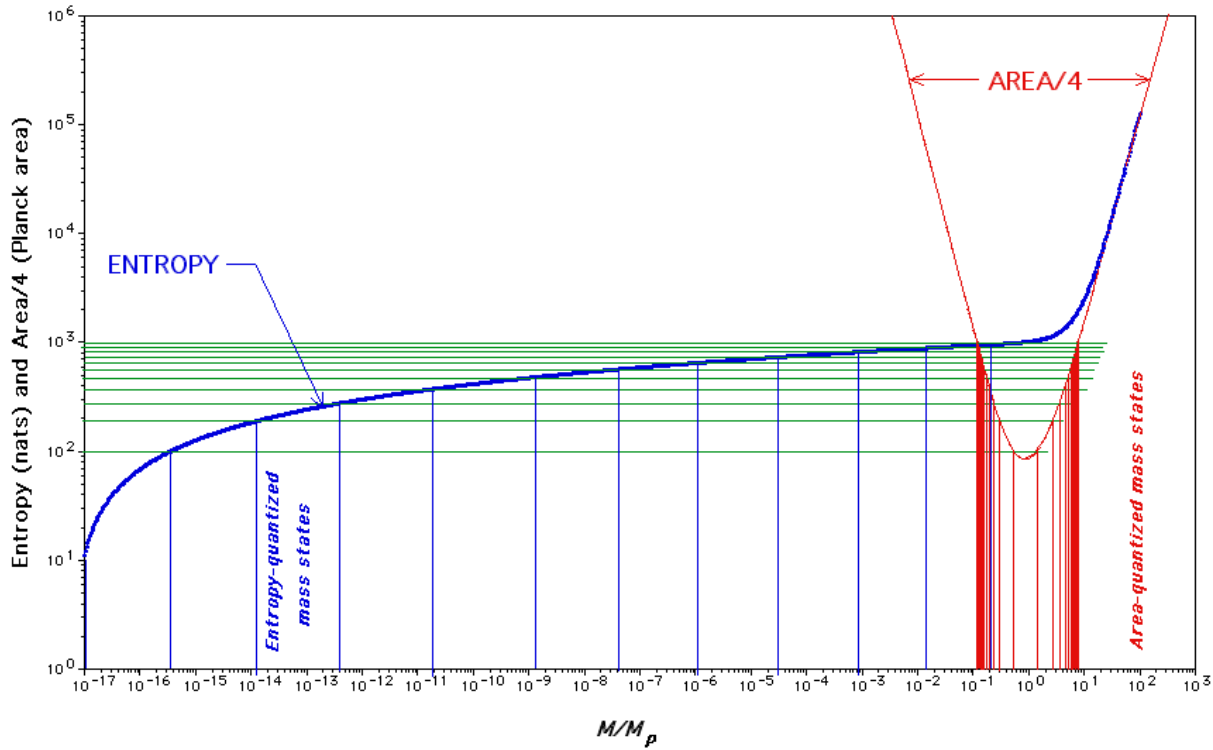

*Fig. 2: A simplified view of the two quantization schemes; for clarity, quantum jumps of $128 A_o$ (or $128 S_0$) have been used.*

Quantization consists of simply taking appropriate equal intervals on the $A/4$ (or $S$) axis, locating the intercept on the curves, and projecting this intercept on the mass axis. For area quantization, there is a natural lowest value of area ($A_{min}$), and there are pairs of states roughly symmetrical (on a log plot) around this point. Quantum numbers are positive above this point, and mass increases as quantum number increases from $n=1$. Below this point, quantum numbers are negative, and mass decreases as quantum number increases negatively from $n=-1$. Even with the large quantum jumps of fig. 2, the area-quantized states become very dense on the mass scale as larger (absolute) quantum numbers are reached.

For entropy-quantization, a casual inspection of fig. 2 shows that above $M_p$, $S \rightarrow A/4$ and the two quantization schemes are indeed equivalent. However, at and below $M_p$ the two schemes are drastically different. The logarithmic form of $S$ requires a minimum mass state $M_L$ in order to have finite and positive entropy, and thus there is a state with quantum number $n = 1$. As $n$ increases, there corresponds a single unique state for each $n$, and the mass spacing of such states is much broader, until the region above $M_P$ where the behavior becomes congruent with the closer-spaced states of area-quantization.

It is implied by fig. 2 that area-quantization in SSGS makes a very odd, if not untenable prediction for the lowest lying scalar states. The states are already getting very compressed even as the mass is reduced to only 0.1 $M_P$; it is natural to ask if, compressed even further over the next 16 decades of mass reduction, and with reasonably smaller quantum jumps, they start to overlap and make a continuum. The same question will have to be asked for entropy quantization, though the low-mass spectra just shown seem better behaved.

*So we must again entertain the conceptual issue of which physical parameter, area or entropy, is to be quantized, and then, with either, use this consistently in both sub- and super-Planckian regions.* Without bias, we do both. To get a better quantitative understanding, we use the asymptotic forms of Appendix A, separating the analysis into sub-Planckian regions (below about 0.01 $M_P$) and super-Planckian regions (above about 100 $M_P$). For this treatment, we use the quanta $A_0$, $S_0$, as defined in (1,3) in both regions. In these asymptotic regions, the solutions for the states are mathematically trivial; we will return to the details of the intermediate region where $M \approx M_P$, where the answers are more subtle, in sec. 4.

*Area-quantization:*

For the super-Planckian region,

(4) $\quad A = 16\pi G_N^2 M^2 / c^4 \quad$ for $M$ sufficiently large, because $G \to G_N$. Thus

(5) $\quad dA = \dfrac{32\pi G_N^2}{c^4} M dM$, and in the limit of large mass, the spacing of levels is

(6) $$\Delta M \to \dfrac{c^4 A_o}{32\pi G_N^2 M} = \dfrac{(\ln 2) M_P^2}{8\pi M} \ .$$

If we take the simplest case where successive states are given by $nA_o$ upward steps in area, relative to the state of least area, $n = 1$, where $A = A_{min}$ (in Planck units):

(7) $$M_n = \dfrac{1}{2}\sqrt{\dfrac{(\ln 2)(|n|-1)}{\pi}} M_P, \quad n = \ldots p, p+1, p+2, \ldots,$$

effectively the same as in the Bekenstein model. The value of $p$ is the first integer for which the asymptotic approximation is valid,[1] and $A_{min}$ is negligible. For large $n$, the spacings derived from (7) show the expected behavior given by (6).

---

[1] This ambiguity will be resolved in Sec. 4.

For the sub-Planckian region,

(8) $$A = 16\pi\hbar^2 / M^2 c^2 \; ; \text{ the corresponding gap spacing is}$$

(9) $$\Delta M \to -\frac{(\ln 2)M^3}{8\pi M_P^2},$$

but now it holds in the limit of small $M$ as the levels become much less than the Planck mass. (The negative sign just shows that successive steps in mass are downwards instead of upwards as in the super-Planckian region). The states corresponding to $|n|A_o$ upward steps in area are then

(10) $$M_n = 2\sqrt{\frac{\pi}{(\ln 2)(|n|-1)}} M_P, \quad n = \ldots -p, -p-1, -p-2, \ldots \; .$$

Once again, for large $|n|$, we get the expected spacing shown in (9). At this point it is interesting to recall that the SSGS model is dual in many parameters, including length (and thus area), under the transformation $M/M_P \to M_P/M$. As a reminder of the duality, we use a notation $(n, -n)$ to denote the dual pair. Thus it is no surprise that $M_{|n|}/M_P \to M_P/M_{-|n|}$ for the two regions.

This area-quantization then shows states that are widely spaced in the vicinity of $M_P$, with gaps about one percent of $M_P$, and which cluster more tightly together as $n$ gets large, either positively or negatively, with the duality just noted. In spite of duality, the clustering is substantially tighter in the negative-$n$ region, a feature not obvious on a log plot; while the positive-$n$ states distribute to infinite $M$, the corresponding dual states ($n < 0$) are limited to $M > 0$. These simple formulae break down in the vicinity of $M_P$, but the qualitative features are shown diagrammatically in fig. 3, shown below after the parallel discussion of entropy quantization.

*Entropy-quantization:*

Now we examine the alternative method for quantization. As mentioned earlier, there is no difference in the two methods for the super-Planckian region; the asymptotic value of $S$ is still given by (2). The states are the same as previously found in (7), and their spacings as found in (6).

For the sub-Planckian region, we would still want the implied quantum of entropy $S_o$ to be universal. The asymptotic form for $S$ in that region, (Appendix A), is

(11) $$S = S_L + 8\pi \ln(M/M_L).$$

In the SSGS, $M_L$ is the mass of the lowest scalar state and $S_L$ is the intrinsic entropy of that state, representing the minimal information contained in a shielding horizon; these are arbitrary parameters of the scenario. Differences in entropy are independent of $S_L$. This form yields the asymptotic values for the gap spacing in entropy-quantization:

$$dS = 8\pi \frac{dM}{M}, \qquad (12)$$

and in the limit of *small M*, the gap becomes

$$\Delta M \to \frac{(\ln 2)}{8\pi} M, \text{ or more precisely}^2, \qquad (13)$$

$$\Delta M \to (e^{\left(\frac{\ln 2}{8\pi}\right)} - 1) M. \qquad (14)$$

The sub-Planckian states are then

$$M_n = M_L e^{\left(\frac{(\ln 2)}{8\pi}\right)(n-1)}, \qquad n = 1, 2, 3, \ldots, \qquad (15)$$

which leads to the gap spacing (14), when $n$ becomes *small*. (Here we take $n = 1$ to be the ground state, so that $M_1 \equiv M_L$ and $n = 1, 2, 3,\ldots$; there are no negative $n$.) These formulae for entropy-quantization also break down in the vicinity of $M_P$, but the qualitative features are as in fig. 3, below, shown with the previous results from area-quantization.

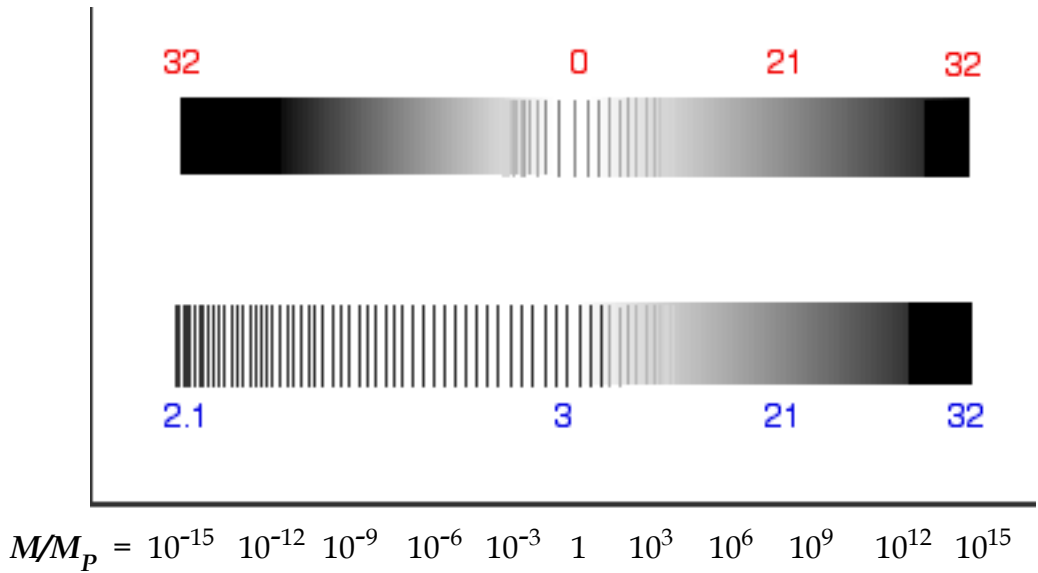

$M/M_P$ = $10^{-15}$ $10^{-12}$ $10^{-9}$ $10^{-6}$ $10^{-3}$ 1 $10^{3}$ $10^{6}$ $10^{9}$ $10^{12}$ $10^{15}$

*Fig. 3: (top) A diagrammatic view of the mass states as determined by area-quantization; (bottom) a similar view of the mass states as determined by entropy-quantization. The numbers near the states indicate the base 10 log of the quantum number $|n|$ for that state and quantization scheme. ( n values to the left of $10^0$ are negative for area-quantization.)*

---

[2] The sub-Planckian region can admit large values of $n$ for which the small-$n$ approximation is invalid; see eq. 25.

A word of explanation concerning this figure is called for. For states one quantum apart in A or $S$, the enormous density of states defies pictorial representation. Fig. 3 uses shading to represent the density, and shows some individual states when feasible, spaced a large (constant) number of basic quantum units apart and represented as lines. The scales of $n$ and $M$ are indicated to illustrate the enormous span of values in this model. Detailed numerical results are deferred to Sec. 4, where the question of whether or not the dense regions are continua will be explored.

The entropy-quantized mass states also cluster for both large and small $n$, so at first glance it appears that that we have a manifestation of states somewhat like in the case of area-quantization. This conclusion is misleading, as is seen by a comparison of gap sizes in (9) and (14): $\Delta M/M \propto (M/M_p)^2$ in the first case (area) and is *constant* in the second (entropy). In entropy-quantization, the index $n$ plays a different, non-dual role and leads to states in the sub-Planckian region, which are quite asymmetric to those in the super-Planckian region.

Before examining these results more quantitatively, we generalize the quantized states to include more than integer steps of $S_o$ or $A_o$. The obvious motivation for this is that the conjectured physical reason for quantization in the SSGS is that each stabilized black hole has a different structure for its horizon, and it is naïve to suppose that this requires (in entropy-quantization) only one additional bit of information per step. For area-quantization the naïve assumption might be more reasonable, but one can generalize similarly anyway. In any case, the parameters for the generalization are defined in fig. 4, where with suitable exchanges between $S$ and A, both types of quantization are generalized. The re-derived formulas for gaps and masses include the generalization wherein $S_L = S_o$ is replaced by $S_L = \eta S_o$, to reflect a possibility that the ground state $n = 1$ has an arbitrary number of entropy quanta needed to define it. Similarly, the steps between states can be an arbitrary number of quanta, defined as $q$. This leads to a formula for the overall entropy of each state:

(16) $$S_n = S_o (\eta + [n-1]q), \text{ and for the quantized area,}$$

(17) $$A_n = A_{min} + A_o (|n|-1)q .$$

In the description of this generalized spectrum, $q$ is assumed to be an integer constant, but in reality could be an integer variable. That is to say, the amount of information (or area) needed to establish the shielding horizon for the next higher state could be complicated and not just the same repeated jump[3]. In such a case the model would have no predictive power, unless there were some distinct regularity to the function $q(n)$.

---

[3] This did not occur in atomic physics: the equivalent for Bohr quantization was $q=1$, $\eta=0$.

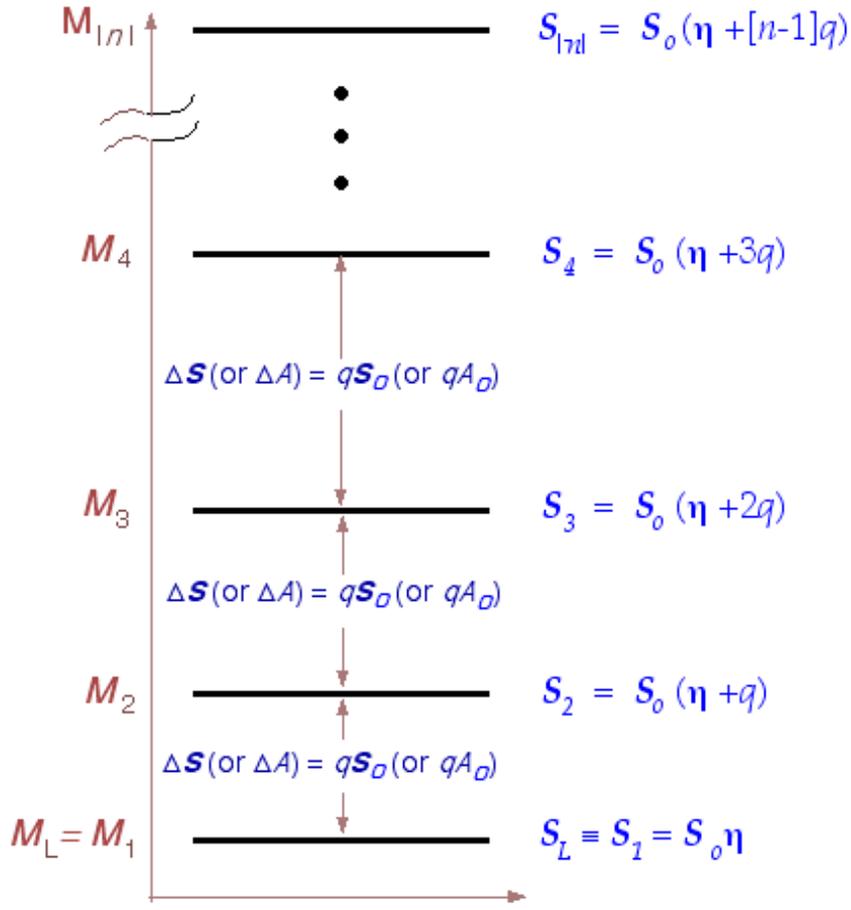

*Fig. 4: The definitions of the quantities q and η in the generalized formulation of mass states; the positions of the masses are generic and not representative of any numerical result. For area-quantization, replace $S_o\eta$ by $A_{min}$.*

The final generalized formula for masses and spacings are given by:

(18) $\quad M_n = \sqrt{\dfrac{(\ln 2)q(|n|-1)}{4\pi}} M_P$

(area-quantization, super-Planckian)

(19) $\quad \Delta M \to \dfrac{(\ln 2)q M_P^2}{8\pi M}$ ;

(20) $\quad M_n = \sqrt{\dfrac{4\pi}{(\ln 2)q(|n|-1)}} M_P$

(area-quantization, sub-Planckian)

(21) $\quad \Delta M \to -\dfrac{(\ln 2)q M^3}{8\pi M_P^2}$ ;

(22) $$M_n = \frac{1}{2}\sqrt{\frac{(\ln 2)q(n-1)}{\pi}}M_P$$

(entropy-quantization, super-Planckian)

(23) $$\Delta M \to \frac{(\ln 2)qM_P^2}{8\pi M} \; ;$$

(24) $$M_n = M_L e^{\left(\frac{(\ln 2)q}{8\pi}\right)(n-1)}$$

(entropy-quantization, sub-Planckian)

(25) $$\Delta M \to (e^{\left(\frac{q(\ln 2)}{8\pi}\right)} - 1)M \; .$$

The masses and splittings are independent of the entropy value $\eta S_o$ given to the lowest state $M_1 = M_L$, as it should be for physical processes that depend only on a change in entropy.

*State widths:*

While the above stratagems have allowed us to define states, we have no formal theory to tell us what the widths of these states should be. At least a crude estimate is needed, because if the gap $\Delta M$ between two states becomes appreciably smaller than their widths, the spectrum is better described as a continuum. The SSGS does not provide us with a time-dependent wave function, which could be Fourier-inverted to yield the width of states. Instead, all the information of SSGS is lodged in the function $M(t)$, and in the quantized values $M_n$. Estimates of the widths of the states can be made in various ways, and we use two very different methods that felicitously arrive at the same answer.

Method 1 (statistical): We interpret $M(t)$ to be a statistical average over a large variety of quantized modes of decay. Using a simple two-state system of masses $M_1$ and $M_2$, what we can physically observe in a single trial (ST) is:

(26) $$M_{ST}(t) = M_2 + (M_1 - M_2)H(t_d - t),$$

where $H(y)$ is the Heaviside function ($H(y) = 1$ if $y > 0$, and $H(y) = 0$ if $y < 0$); $t_d$ is the time at which $M_1$ decays into $M_2$. Averaging $M_{ST}$ over the decay probability distribution for $M_1$ (with lifetime $\tau_1$)

(27) $$P(t_d) = \frac{1}{\tau_1}\exp\left(-\frac{t_d}{\tau_1}\right), \text{ we find the statistically-averaged observable:}$$

(28) $$M(t) = M_2 + (M_1 - M_2)e^{-t/\tau_1},$$

and by differentiating and manipulating the terms we arrive at the result:

$$\text{(29)} \qquad \frac{1}{\tau_1} = -\frac{dM}{dt}\bigg|_{t_1} \frac{1}{M(t_1)-M_2} = -\frac{dM}{dt}\bigg|_{M_1} \frac{1}{M_1-M_2} \;.$$

Method 2 (heuristic): The previous method does not include multiple states with varying lifetimes. If we instead interpret $\tau$ as a mean transition time from a given state to the next lowest state, and assume that $dM/dt$ reflects the classical limit as the mean value of radiation power (black hole energy loss per unit time), then $dM/dt = -(M_n-M_{n'})/\tau$ for two closely-lying states, resulting in the same form as method 1, for two closely-adjacent states. This seems correct in another sense: What limits the length of an interval $\delta t$, in which we sample a mass described by $M(t)$? We can make the interval arbitrarily short and artificially induce a large width, but is there any *upper* bound to the interval? We know that the interval $\delta t$ samples a mass range $\delta M = -(dM/dt)\,\delta t$. For the measurement of the mass to be pertinent to the mass $M_n$, $\delta M$ must be smaller than the gap $\Delta M$ between quantized states. This line of argument gives an estimator for the inverse transition time:

$$\text{(30)} \qquad \frac{1}{\tau} \equiv \frac{1}{\delta t} \approx -\frac{dM}{dt}\bigg|_{t_1} \frac{1}{\Delta M_n}, \quad \text{and thus } \Gamma \approx \frac{\hbar}{c^2 \tau} \;.$$

This constitutes a lower limit for $\Gamma$, which in principle could be attained, suggesting it is a good estimate available for the natural width.

How could we arrive at such an answer with no detailed knowledge of the partial widths of the original state? In this case, those would reflect all the combinations of particle emission modes into which the black hole could evaporate. In the classical formula for emission power, $dM/dt$, the effects of these multiple modes are already incorporated into the generalized Stefan-Boltzman constant $\alpha$, which includes all known particle emission modes as well as spin-dependent gray-body factors [15]. Thus it is plausible that $M(t)$ represents, via its thermodynamic origins, a statistical average over all such decay modes (not just the two-state system we used above). Knowing $\Delta M_n$ from a broad general principle—i.e., the variable, $S$ or $A$, is an adiabatic invariant—completes the knowledge necessary to describe the basic physical system.

The resulting transition times are displayed in fig. 5. We stress that the functions plotted depend not only on the method of quantization, but in addition on the parameters of the quantization: $M_L$, $q$, $\eta$ and $A_0$ (or $S_0$). The curves are shown separately for entropy and area quantization, but both use a "standard" set of parameters $q = \eta = 16$, $M_L = 100$ GeV, and $A_0$ (or $S_0$) as described. Also note that the mass-dependence of $\alpha$ in a realistic model will modify not only the time scale of $dM/dt$, as noted in [1], but can also modify transition times and widths. Once the mass spectra are actually known, the SSGS model can be reworked to reflect these changes. While we expect the same general picture to hold true, the details will certainly be modified.

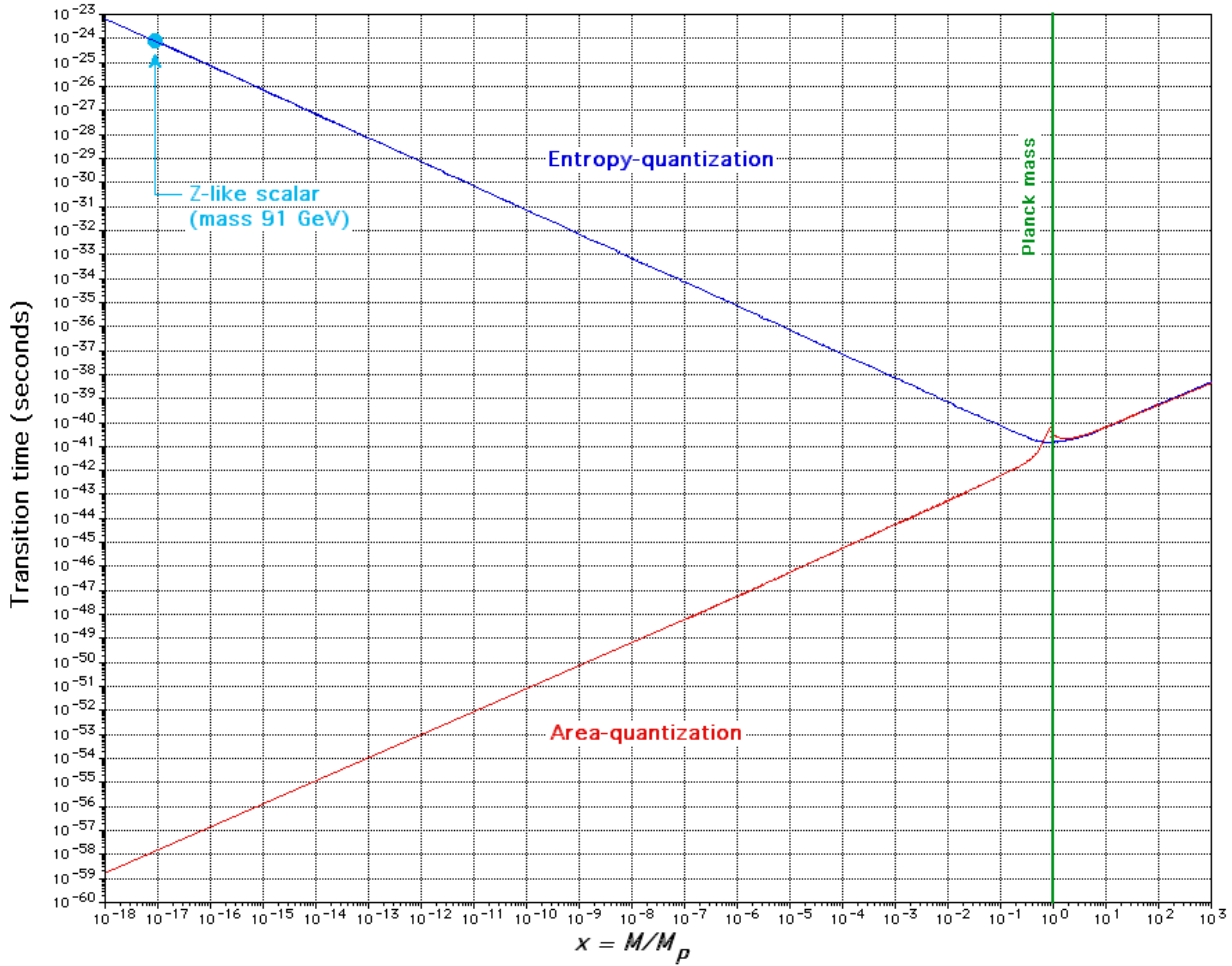

*Fig. 5: Estimation of transition times of mass states in SSGS for the two quantization schemes.*

Are these transition times reliable?  We can inspect the super-Planckian region, where we think we know how classical black holes behave; in this region, entropy- and area-quantization give results identical to one another.  The classical Hawking *entire* life of a black hole is given by $T = M^3/3\alpha$.  The entire life $T$ is not a pertinent quantity if one is considering a mass measurement of a particular quantum state, as discussed earlier.  If those considerations are included, along with the appropriate $\Delta M_n$, then we find the equivalent Hawking $\tau \propto M$, which is also the result of fig. 5.  Ultimately, though, there is no real test of (30), because no experiments are yet capable of detecting spectral lines from readily available astrophysical black holes.  Our result, however, is algebraically identical to a more sophisticated treatment of Bekenstein and Mukhanov[4], and is used consistently in both super- and sub-Planckian regions.

In the sub-Planckian region, the only basis for comparison is to look for lifetimes that are similar to those of elementary particles of similar masses; this, too, is of somewhat limited utility because no scalars have ever been found, and the results are likely spin-dependent.  Fig. 5 shows that there is a gigantic difference in the predictions of the two types of quantization for black-hole particle transition times.  For instance, at about the TeV mass scale, area-quantization predicts a particle lifetime of order $10^{-57}$ s, and one even shorter at the GeV mass scale($10^{-60}$ s). This clearly has no connection with any experimental physical reality, and of course leads to continuum behavior (to be discussed below).

Entropy-quantization in the sub-Planckian region fares much better, with transition times looking entirely consistent with low-energy experience. At TeV masses, the predicted transition time is of order $10^{-25}$ s, and at a GeV, $10^{-22}$ s. The particle most similar to our uncharged scalars might be the $Z^0$ (neglecting its vector nature), and the prediction shown in fig. 5 is for a transition time corresponding to a width of 0.8 GeV. The experimental width is $\Gamma_Z \approx 2.5$ GeV [16], giving some credibility to our transition time calculations. Conversely, if one believes the estimation is correct, the above set of differences in transition times provides a strong discriminant between entropy- and area-quantization schemes.

The next section uses all of the above derivations to quantify numerically both the spectra and expected decay widths.

**Section 4. Numerical results and discussion**

Now we are prepared to answer the question as to which quantization scheme, if either, is preferred, given that they are not equivalent for sub-Planckian masses. The answer hinges in part on whether a given scheme yields discrete non-overlapping states, or produces a continuum of unresolvable states. We must now be precisely quantitative—the schematic diagrams are not sufficient. First, we pick specific values of $q$ and $\eta$ to make quantitative predictions. For this example, we choose $q = \eta = 16$: that says that the physics requires a change of 16 quanta of area $A_0$ to move from one mass state to another. With entropy-quantization, that also corresponds to 16 quanta of size $A_0/4 = \ln 2$, or one 16 bit word of information.[4] We then use analytic formulae (Appendix A) for both $A(M)$ and $S(M)$, and numerically find their inverse functions $M(A)=A^{-1}\{A\}$ and $M(S)=S^{-1}\{S\}$ to solve the equations

$$M_{|n| \text{ or } -|n|} = A^{-1}\{A_{min} + (|n|-1)qA_o\} \quad , \quad |n| = 1, 2, 3, \ldots \tag{31}$$

$$M_n = S^{-1}\{S_o[\eta + (n-1)q]\}, \quad n = 1, 2, 3, \ldots \tag{32}$$

for the quantization of area and entropy, respectively.

The results are shown in the rather busy figs. 6a, 6b, where we have plotted the dimensionless $\Delta M/M_P$ for the two quantization schemes, plus the previously derived estimate for the widths of states at relative mass $x = M/M_P$. The $n$-value for any quantized mass is also shown for both schemes; one selects integer $n$ from the continuous curves.

---

[4] If this choice seems overly whimsical, rest assured that the conclusions hold very similarly for q>1 and any η. It is only a question of the scale taken on by the quantum index $n$.

Page 13

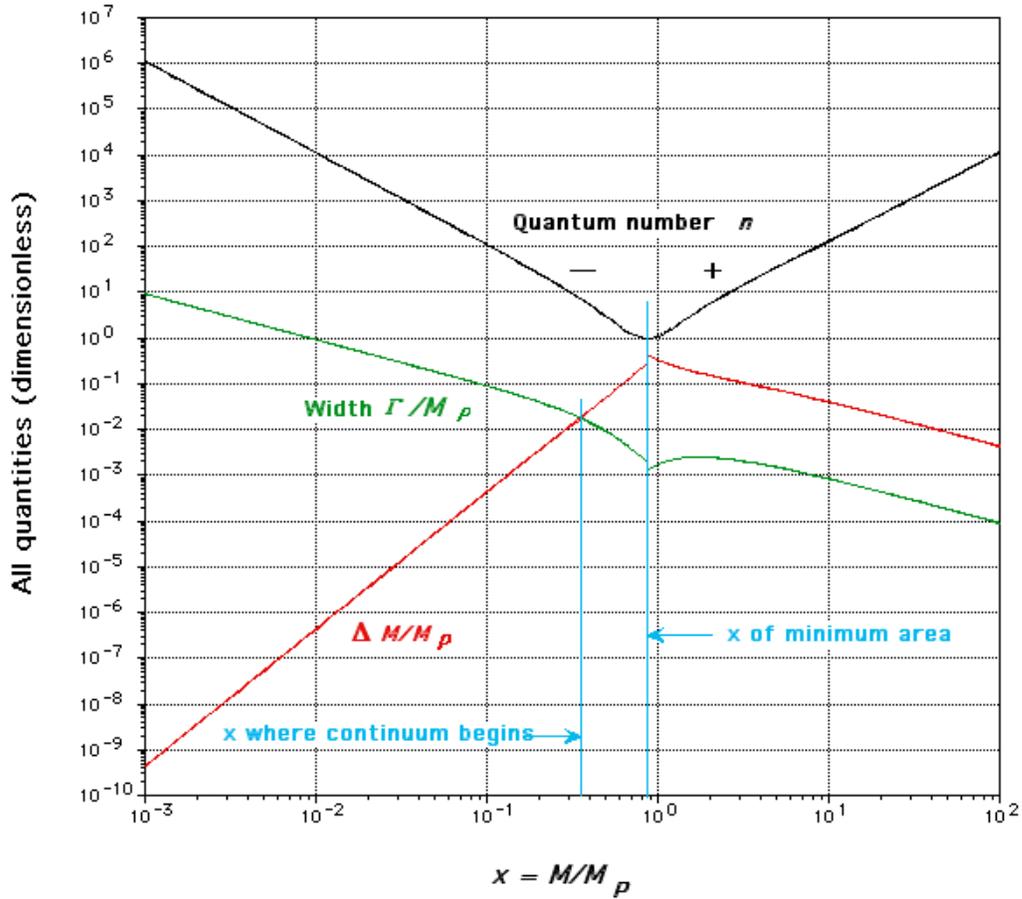

*Fig. 6a: Distinguishability of individual mass quantum states in area-quantization; the discontinuities near x =1 are artifacts of the asymmetry of A(x) and the integral nature of n.*

Now it is easy to deduce where states will merge and where they will be distinguishable. The condition is simply that a particular $\Delta M/M_P$ curve must be above the $\Gamma/M_P$ curve. For area-quantization, this is satisfied only if $x = M/M_P > \sim 0.4$. For the choice of $q$ made above, that means only for quantum numbers $-6 < n < +\infty$. Physics at the TeV level would correspond to $n \approx -10^{32}$, where $\Gamma/\Delta M \approx 10^{33} \text{GeV}/4.4 \cdot 10^{-30} \text{ GeV} \approx \mathcal{O}(10^{62})$, a rather convincing continuum! It isn't clear to these authors how this continuum almost up to $M_P$ would manifest itself in S-wave amplitudes in a cross section, but it would certainly say that no isolated scalar particle would be found until an energy scale where the widths were small compared to the gaps. This scenario shows little experimental promise at best, and at worst is simply inconsistent with a total absence of scalars below $\sim 100$ Gev/c², and with experimentally sensible lifetimes. It does say, as expected from our mimicking Bekenstein's work, that in the super-Planckian region, black holes would form semi-stable states, with, e.g., $\sim 1$ TeV transitions occurring at $M \sim 5.3 \cdot 10^{34}$ Gev ($\sim 10^8$ kg), $n \sim 2 \cdot 10^{31}$ and $\Gamma \sim 0.02$ TeV. Still, since this scenario appears unredeemable in the sub-Planckian region, we pursue it no further in this paper.

In fig. 6b, we find that a parallel examination reveals a quite different prospect for the scenario where we quantize entropy. It gives the same plausible result just discussed for the super-Planckian region, where it is indistinguishable from the classic theory (except for the labeling of the index *n*). But in the sub-Planckian region, because of the logarithmic behavior of the entropy there, the quantum states are stretched out. Fig. 6b reflects this, in that it shows a radically different behavior for *n(x)*, and because $\Delta M/M_P$ (entropy) is *always* well above the $\Gamma/M_P$ curve. The quantized mass states do not form a continuum anywhere (at least not from intrinsic width effects), below or above the Planck mass. In fact, this analysis predicts that the widths of the uncharged neutral scalar states, for low-lying entropy-quantized masses, are a constant 0.74% of the observed mass; the gaps between states are ~56% of the observed mass. Thus $\Gamma$ is about 1.3% of the gap size $\Delta M$ and there is no continuum region. This holds all the way down to the lowest state at $M_L$, required for self-consistency of the model.

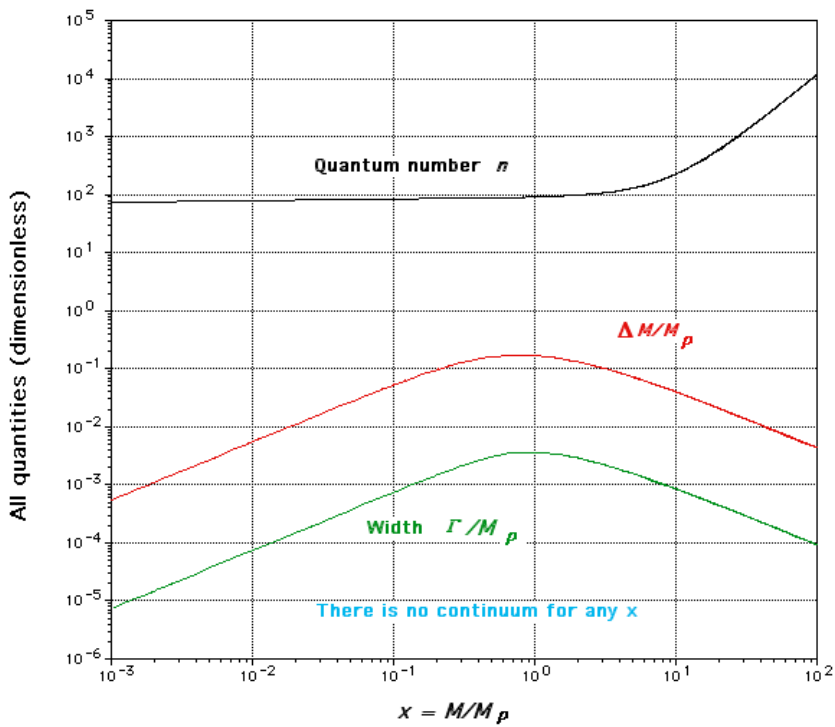

*Fig.6b: Distinguishability of mass quantum states in entropy-quantization.*

Given this encouraging picture for the entropy-quantized scheme, we offer some predictions for CERN. Fig. 7 shows individual low-lying states for the parameters given, drawn with their calculated $\Gamma$'s when such $\Gamma$'s are greater than the minimum line width available for the drawing. For completeness, the behavior of the entropy-quantized states near the Planck mass is also shown in fig. 8. Only a few characteristic *n* are shown to simplify the diagram. Note that there are "only" 91 states from 100 GeV up to the Planck mass for this particular parameterization. While that may seem like a plentitude, high-energy physics has explored only about $10^{-17}$ of the possible spectrum up to $M_P$. The spectral density of particles already found is enormous compared to what this example predicts for higher energy regions. To future experimentalists, those regions may indeed seem like a desert.

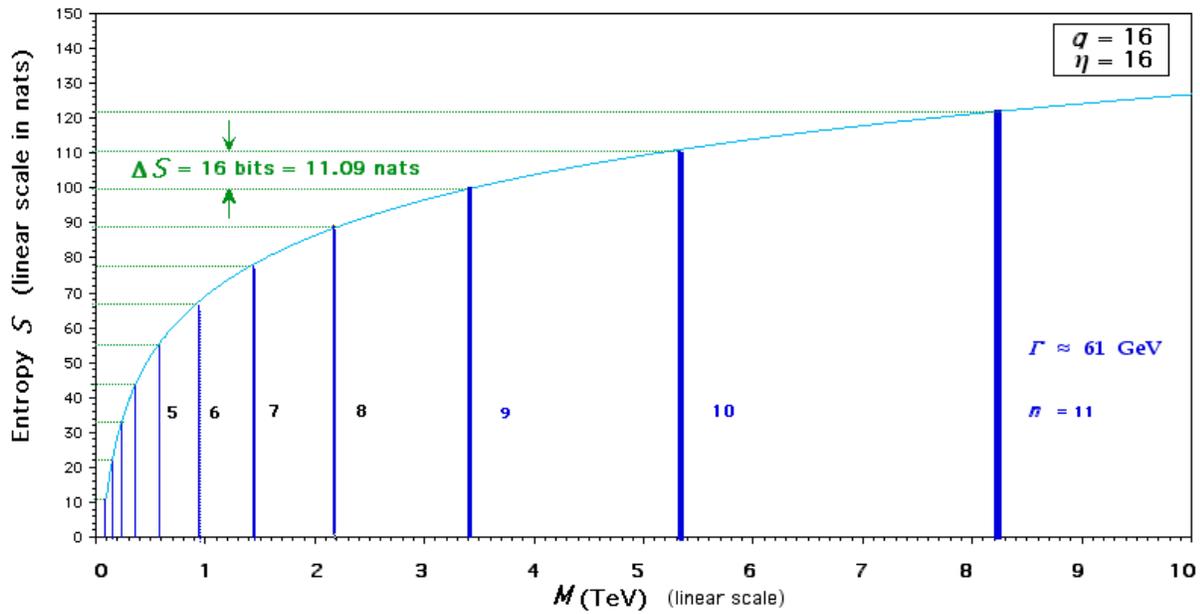

*Fig. 7: Entropy quantization for the case $\eta = 16$, $q = 16$. Mass states are shown at absolute masses and (where visible) with their predicted line widths.*

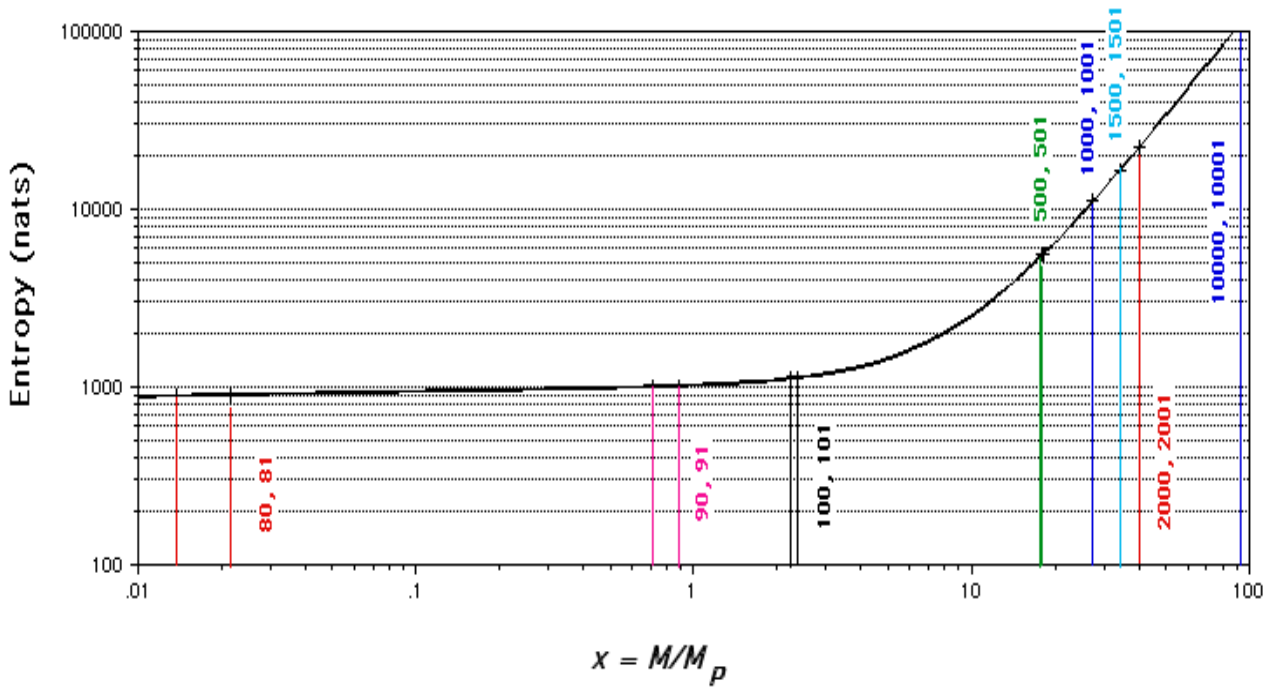

*Fig. 8: Mass state separation in the vicinity of $x = 1$ for entropy-quantization similar to Fig. 7; numbers by lines designate pairs with adjacent index n. Widths of states are negligible relative to gaps, even as states seem to merge on the diagram.*

Finally, Table 1 below shows the array of lowest-lying accessible states of uncharged scalars to be expected at CERN, if $M_L$ has been correctly guessed at 100 GeV and the required number of fundamental quanta per state is 16. The usual but arbitrary selection $\eta = 16$ is used, indicating that the ground state also requires 16 bits of information. We stress that this is an example of the *pattern* to be expected, rather than some particular masses. *The real prediction of the SSGS is the exponentially increasing pattern of masses stemming from the logarithmic dependence of entropy.*

| $n$ ($q=16$) ($\eta=16$) | $n'$ ($q=1$) ($\eta=16$) | $M_n$ (GeV) | $M_{n+1} - M_n$ (GeV) | $\Gamma_{SSGS}$ (GeV) | Entropy (includes $S_L$) (bits) | Entropy (includes $S_L$) (nats) |
|---|---|---|---|---|---|---|
| 1 | 1 | 100 | 55 | 0.74 | 16 | 11.1 |
| 2 | 17 | 155 | 87 | 1.14 | 32 | 22.2 |
| 3 | 33 | 242 | 131 | 1.78 | 48 | 33.3 |
| 4 | 49 | 376 | 208 | 2.77 | 64 | 44.4 |
| 5 | 65 | 584 | 324 | 4.30 | 80 | 55.5 |
| 6 | 81 | 908 | 504 | 6.68 | 96 | 66.5 |
| 7 | 97 | 1412 | 783 | 10.4 | 112 | 77.6 |
| 8 | 113 | 2195 | 1218 | 16.2 | 128 | 88.7 |
| 9 | 129 | 3413 | 1893 | 25.2 | 144 | 99.8 |
| 10 | 145 | 5306 | 2943 | 39.1 | 160 | 110.9 |
| 11 | 161 | 8249 | 4576 | 60.7 | 176 | 122.0 |
| 12 | 177 | 12825 | 7114 | 94.2 | 192 | 133.1 |

*Table 1: The lowest-lying states of neutral scalar black holes, in an example of the entropy-quantization scheme of the SSGS model. If the states are fine-grained with only one bit per level, n' shows the corresponding quantum number.*

For any $M_L$ and any $q$, the generalized eqs. (24, 25) tell us, given the first two states $M_1 \equiv M_L$ and $M_2$ from experiment, where the rest of the expanding tower of particles would lie. The spacings will look linear in the mass only for a few lowest-lying states, because both the mass levels and their spacings depend *exponentially* on $n$. The SSGS model thus intrinsically admits a mechanism suggesting that elementary particles can show gigantic differences in mass. In SSGS, the quantity $q$ is the number of fundamental bits of information needed to specify the next higher shielding horizon. If $q=1$ and only one bit is needed (which seems unlikely), then there are ~1500 uncharged scalar states between 100 GeV and the Planck mass; if $q=100$, there are only ~15. In this latter case, if $M_L$ = 100 Gev, $M_2$ would be at ~1.6 TeV and the pattern of higher states would be less accessible at LHC in 2007. The main example illustrated in the table, $q = 16$, is more felicitous for significant testing at LHC.

*Simultaneous quantization of area and entropy*

Area-quantization, dismissed here as unrealistic because of its low mass continuum and unreasonable lifetimes, has become almost a sacred mantra of quantum gravity. Can we save it in the SSGS model? If *simultaneous* quantization of both entropy and area is required, then area-quantization does *not* lead to any low-mass continuum problems; the low-mass entropy-quantized eigenvalues $_SM_n$ are always matched (to well within their widths) by the very dense values of the area-quantized eigenvalues $_AM_{n'}$. Then the requirement of a match dilutes the high density of $_AM_{n'}$. By eliminating the high density, it also gets rid of the overly-short lifetimes. There is also no problem in getting matches well above $M_P$, where the two schemes eventually converge. But this still leaves a region where, depending on the values of the parameters $M_{P'}$, $q$, $\eta$, $S_o$ $A_{min}$, $A_{o'}$, there may be many states not normally congruent in the two schemes. This in turn puts that many constraints on the parameters, in order to obtain a match. (Of course, an alternative might be that there would simply be no states in this region). An optimist might view this as wonderful—a method by which all parameters would be determined. We suspect that such a highly over-constrained system would have no "natural" solution, and a number of matching states exactly equal to the number of parameters would be obtained. We have not attempted a fit to this parameterization.

Instead, we suggest that in SSGS, area is not an adiabatic invariant near and below the Planck mass, because the system is in strong transition as it undergoes the "phase transition" there. Instead, the truly adiabatic invariant is entropy, as the plots show. Well above the Planck mass, $S \propto A$ and the traditional picture is restored.

**Concluding remark**

In SSGS, we attempted to adopt a Bohr-like approach to obtain insight into how small black holes would behave if gravity is strong but shielded. We had hoped to quantize the scalars and thus be prepared to compare them with real discoveries of such particles. In that work, we failed to quantize any states, a shortcoming possibly now resolved in this paper. We remain at a disadvantage relative to a Bohr-like approach (in at *least* the following respect!) because the spectrum of putative scalar states is not yet known experimentally, either to set the parameters of our model or to throw it out. That turns into an advantage if and only if the eventual results at least generally confirm our predictions for the morphology of states.

**Acknowledgements**

We thank Fred Kuttner and Bruce Rosenblum for their critical readings of this paper, Melanie Mayer for her attention to our prose and Benjamin Lubin for his technical advice in the preparation of the manuscripts.

## Appendix A: Exact equations and asymptotic forms from [1]

In SSGS, the black hole evaporation differential equation takes the form:

$$(33) \qquad \frac{dM}{dt} = \frac{-\alpha}{M_P^2}\left(\frac{4x}{\left(1+\sqrt{1+4x^2}\right)^2}\right)^2,$$

where $x$ is defined as $x = M/\mu M_P$. The black hole evolution time as a function of mass (the inverse function of $M(t)$) is then given by

$$(34) \qquad t - t_0 = -\mu^3 \frac{M_P^3}{16\alpha}\left\{\begin{array}{l} \frac{16}{3}(x^3 - x_0^3) + 32(x - x_0) - 8\left(\frac{1}{x} - \frac{1}{x_0}\right) + 8\sqrt{4 + \frac{1}{x^2}}(x^2 - 1) \\ -8\sqrt{4 + \frac{1}{x_0^2}}(x_0^2 - 1) + 20\ln\left(\frac{x}{2}\sqrt{4 + \frac{1}{x^2}} + x\right) - 20\ln\left(\frac{x_0}{2}\sqrt{4 + \frac{1}{x_0^2}} + x_0\right) \end{array}\right\},$$

where $\alpha = 7.8 \cdot 10^{26}$ g$^3$/s.

In the current paper, $\mu$ is always unity, and is so in all of the following equations.

$$(35) \qquad \frac{G}{G_N} \equiv g = \frac{1}{4}\left\{\frac{1}{x} + \sqrt{\frac{1}{x^2} + 4}\right\}^2$$

$$(36) \qquad A = 16\pi(gx)^2, \text{ in Planck units } \frac{\hbar G_N}{c^3}$$

$$(37) \qquad S = S_L + 4\pi\left\{2\ln\left(\frac{x}{x_L}\right) + (x^2 - x_L^2) + \sqrt{4x^2 + 1} - \sqrt{4x_L^2 + 1} + \ln\left[\frac{1 + \sqrt{4x_L^2 + 1}}{1 + \sqrt{4x^2 + 1}}\right]\right\}$$

Asymptotic forms, super-Planckian region:

$$(38) \qquad t - t_0 = -\frac{M_P^3}{3\alpha}(x^3 - x_0^3) \qquad \text{(The Hawking formula)}$$

$$(39) \qquad \frac{G}{G_N} \equiv g = 1$$

$$(40) \qquad A = 16\pi x^2, \text{ in Planck units } \frac{\hbar G_N}{c^3}$$

$$(41) \qquad S = 4\pi x^2$$

Asymptotic forms, sub-Planckian region:

(42) $$t - t_0 = \frac{M_P^3}{\alpha} \frac{1}{x}$$

(43) $$\frac{G}{G_N} \equiv g = \left(\frac{1}{x}\right)^2$$

(44) $$A = 16\pi \left(\frac{1}{x}\right)^2, \text{ in Planck units } \frac{\hbar G_N}{c^3}$$

(45) $$S = S_L + 8\pi \ln\left(\frac{x}{x_L}\right)$$